# The coherence of wave-packet-tunable photons

YA LI [1,*], WANRU WANG[1], QIZHOU WU[1], YOUXING CHEN[1,†], CAN SUN[2], HAI WANG[2], WEIZHE QIAO[3]

[1]*School of Information and Communication Engineering, North University of China, Taiyuan 030051, China*
[2]*The State Key Laboratory of Quantum Optics and Quantum Optics Devices, Institute of Opto-Electronics,Collaborative Innovation Center of Extreme Optics, Shanxi University, Taiyuan 030006, China*
[3]*Shanxi Dazhong electronic information Industry Group Co., LTD, Taiyuan 030024, China*
** liya3642@163.com*

**Abstract:** The wave-packet-tunable photons [Optics Express 30, 2792-2802 (2022)] generated by spontaneous Raman scattering (SRS) based on atomic ensemble lay a foundation for the hybrid quantum network to successfully connect quantum nodes with different bandwidths, but the coherence time of wave-packet photons becomes the key factor limiting the distance of entanglement distribution. The coherence of photons deteriorates with the propagation distance of the entanglement distribution. So far, the coherence of wave-packet-tunable photons entangled with an atomic memory has remained unexplored. An unequal arm fiber interferometer is constructed to measure the interference visibility of 150 ns-1.06 μs pulse width wave-packet-tunable photons. The coherence time and bandwidth of the photons can be directly derived from the decay of the visibility in the interferogram as the wave packet photon length increases. The measured results show that the coherence time of write laser is 2.36 μs and bandwidth is 78 KHz, which interact onto atoms can generate Stokes photons with the coherence time is 1.14μs and bandwidth is 156 KHz. The measurement of coherence of wave-packet-tunable photons lays the foundation for establishing the distribution of entanglement between spatially separated memories in hybrid quantum networks, and for establishing a baseline telescope of arbitrary length through wave-packet-tunable photon interference.

## 1. Introduction

Large-scale quantum networks [1-3] and long-distance quantum communication [3-5] rely on quantum Repeaters (QRs) [6] for long-distance entangled distribution. Based on Spontaneous Raman Scattering (SRS) of cold atom ensemble, the atom-photon quantum entanglement has been generated [7-12]. Quantum entanglement experiments have also been experimentally prepared in many matters, such as ion, quantum dots and other different quantum systems [13-21]. Riedmatten's group [22] demonstrated a heralded single photon source with controllable read pulse duration emission time based on a cold atomic ensemble, which can generate photons with temporal durations varying over three orders, which can used for other different quantum systems. Almendros [23] et al. generated a tunable single photon source with coherent time ranging from 70ns to 1.6μs between two remote ion traps. G. B. Xavier et al. [24] experimentally demonstrated stable and adjustable single-photon interference in a 1 km long Mach-Zehnder fiber interferometer, obtaining a single-photon net visibility of 0.97. Seok-Beom Cho et al. [25] reported a single-photon interference experiment in a Michelson-type interferometer composed of two 6-km-long optical fiber spools, which obtain a visibility of more than 98% without subtracting noise counts. Pan Jianwei's team [18] achieved 50km long-range single-photon interference by using single-photon interference scheme, which is basically capable of realizing entanglement between long-distance spatial separated memories. F. Jelezko et al. [26] present a methodology that allows recording the coherence length of photons emitted by a single quantum system in a nitrogen-vacancy (N-V) defect center in diamond and point out that the bandwidth of a single photon can be directly derived from the attenuation of the visibility in the interferogram. The above study of single-photon coherence lays the foundation for the long-distance entanglement distribution, while the single-photon interference scheme also has a higher entanglement yield, but requires a stable phase environment. By designing and implementing a dual phase locking scheme, Pan's [18] team successfully controlled the optical path difference caused by 50 km fiber transmission to about 50 nm. In addition, Daniel Gottesman et al. [27] proposed a scheme to use quantum repeater to build space interference telescopes, aiming to eliminate the resolution limitations of traditional optical interferometers. The long coherence time of the wave packet photons allows in principle for interferometers with arbitrarily long baselines. The technique of generating wave-packet-tunable photons [22, 23, 28] also provides a support for the interface connection of hybrid networks, while the bandwidth of the interface photons from those quantum systems become the key factor for successful long-distance connection of quantum nodes with different bandwidth [29]. However, experimental studies of coherent properties of wave packet tunable photons based on atomic ensembles have not been reported.

We generate wave-packet-tunable photons by varying light-atom interaction duration time in cold atoms [28], then inject those photons into the unequal arm interferometer with different delay fiber to study the coherence properties of wave-packet-tunable photons, when the "head" from the long arm and "tail" from the short arm of packet-tunable photons coincidence to produce an interferogram. The maximum interference visibility of the current state is calculated by modulating the phase when the interference is strongest, and the bandwidth of the photons can be deduced directly from the decay of the maximum visibility in the interferogram as the wave packet length increases [26]. The measurement of the coherence of wave-packet-tunable photons sets the stage for establishing the distribution of entanglement between spatially separated memories in hybrid quantum networks, and for building a baseline telescope of arbitrary length through wave-packet-tunable photon interference.

## 2. Methods

In Ref. [28], we use $^{87}$Rb atomic ensemble as the storage medium for generating wave-packet-tunable photons. As shown in Fig.1, the $^{87}$Rb atomic ensemble is prepared to the ground state at the initial time. The write laser comes from a TOPTICA DLC Pro laser is locked on the saturation absorption D1 line of the rubidium atom to transition with a blue-detuned △ =20MHz. A wave packet-tunable Stokes photon based on SRS can be generated by applying the FPGA to control the write pluses interact with atoms with a tunable duration. Based on the level structure shown in Fig. 1, we know that the bandwidth of the Stokes photon is closely related to the bandwidth of the write laser and the atom. Under different reference (absorption D1 line or optical cavity) mode locking conditions, the bandwidth of write laser is different. We first measure the laser bandwidth using an unequal arm interferometer. Based on the classical write laser measurements described above, we measure the interference visibility of photons.

The bandwidth of the write laser is dominated by low phase noise. Coherence can characterize the intensity of noise in the laser light field. To measure the coherence properties of the photons, we construct an unequal arm fiber interferometer as shown in Fig. 1, which can convert the phase noise of the laser light into intensity noise by means of a delay fiber. The write laser locked to the saturation absorption line of the rubidium atom has stochastic phase noise, so measuring the visibility of the write laser requires the calculation of an average value over the observation time. In order to measure the coherence of the Stokes photons, it is necessary to count the photons in the interference region.

The single longitudinal mode write laser is a monochromatic optical field with a stable amplitude but a perturbed phase. The light field can be expressed as: $E(t)=E_0 \cdot \exp\{i[\omega_0 t+\Delta\varphi(t)]\}$, where $E_0$ is the amplitude of the light, $\omega_0$ is the central frequency of the light field, random phase disturbance $\Delta\varphi(t)=kL(t)$, $k$ is wave vector, $L(t)$ is the length of an optical path due to time-dependent temperature, the output light field of the short arm $L_1(t)$ of the interferometer is: $E_1\exp[i(\omega_0 t+k_0 L_1(t))]$, and the output light field of the long arm $L_2(t)$ is: $E_2\exp[i((\omega_0+\Delta\omega)t+(k_0+\Delta k)L_2(t))]$, where $\Delta\omega$ is the difference of central frequency and $\Delta k$ is the difference of wave vector between $L_1(t)$ and $L_2(t)$. The interference light intensity of the output signal light and the lock light of the unbalanced optical fiber interferometer can be expressed as:

$$\begin{aligned} I_S(t)&=E_1^2+E_2^2+2E_1E_2\cos(\varphi_S(t)) \\ I_L(t)&=E_1^2+E_2^2+2E_1E_2\cos(\varphi_L(t)) \end{aligned} \tag{1}$$

where $\varphi_S(t)=\Delta\omega t+k_S\Delta L(t)+\Delta kL_2(t)$ ( $\varphi_L(t)=\Delta\omega t+k_L\Delta L(t)+\Delta kL_2(t)$ ) is the phase difference of write (lock) light, $\Delta L(t)=L_2(t)-L_1(t)$ is the difference length of interferometer arm, $k_S=\frac{\omega_S}{c}\left(k_L=\frac{\omega_L}{c}\right)$ is the wave vector of write (lock) light, $\omega_S(\omega_L)$ is the frequency of write (lock) light, $c$ is velocity of light. The instantaneous value of intensity $I_S(t)(I_L(t))$ is given by a particular phase fluctuation $\varphi_S(t)(\varphi_L(t))$.

The phase difference between the write and lock lasers can be expressed as: $\Delta\varphi(t)=\varphi_S(t)-\varphi_L(t)=(k_S-k_L)\cdot\Delta L(t)=\Delta k\cdot\Delta L(t)$, which can be controlled by changing the length of delay fiber $\Delta L(t)$ by using the fiber stretcher and the frequency difference $\Delta k$ of the lock and write laser. Due to a small random variable $\delta\Delta L(t)$ in fiber length introduced by the temperature drift and mechanical vibration, meanwhile a variation of the wave vector $\delta\Delta k(t)$ introduced by the laser frequency drift, the phase difference of write (lock) light can be rewritten as:

$$\begin{aligned} &\varphi_S(t)=\Delta\omega t+k_S(\Delta L(t)+\delta\Delta L(t))+(\Delta k+\delta\Delta k(t))L_2(t) \\ &(\varphi_L(t)=\Delta\omega t+k_L(\Delta L(t)+\delta\Delta L(t))+(\Delta k+\delta\Delta k(t))L_2(t)) \end{aligned} \tag{2}$$

and the phase difference between write and lock laser can be rewritten as:

$$\Delta\varphi(t)=(\Delta k+\delta\Delta k(t))(\Delta L+\delta\Delta L(t)) \tag{3}$$

The distribution probability of phase fluctuations $\varphi$ can be expressed as[30] : $P(\varphi)=\frac{1}{\delta\sqrt{2\pi}}e^{-\varphi^2/2\delta^2}$, where $\delta$ is the width of the Gaussian distribution. To obtain some characteristic phase-noise independent value of intensity, which need to average the intensity. Thus we find a direct relation between the visibility and the distribution of the phase noise described by[30] :

$$V(t)=e^{\frac{-\delta^2 t^2}{2}} \tag{4}$$

where $\delta=2\pi\upsilon$ , $\upsilon$ is the bandwidth, *t* is the delay time caused by the difference in length of interferometer arm $\Delta L(t)$.

## 3. Experiment Setup

The interferometer arms use flange joint two 50/50 optical splitter, short arm set a three-loop polarization controller (PC) to control the polarization, long arm set a fiber stretcher (FST) and different length delayed fiber. In the scan model, 100Hz triangular wave signal is injected into high voltage amplifier (HV) to drive the fiber stretcher to periodicity varies with a velocity of 0.14μm/V, thus generating interferogram, as shown in Fig. 2. In the lock model, the interference signal is mixed with the demodulation signal from the electro-optic modulator(AOM), then sent to proportional-integral-derivative (PID) by high-pass (HP) filtering to generate monitor signal, which is feed back to the HV to control FST, thus the length of long arm can make $\varphi_L(t)$ as an even multiple of π, while the lock laser $\varphi_L(t)$ will always be locked at the highest position as shown in Fig.3. The phase difference $\Delta\varphi(t)$ between the write laser and the lock laser can be adjusted by changing the frequency of the lock laser through the AOM. When $\Delta\varphi(t)=2n\pi$, the phase difference is in-phase, where $n$ is a positive integer, the write signal of constructive interference reaches a maximum $I_{max}(t)$; when $\Delta\varphi(t)=(2n+1)\pi$, the phase difference is antiphase, the write signal of destructive interference reaches a minimum $I_{min}(t)$. Thus, the interference visibility can be calculated by the following equation [26]:

$$V(t)=\frac{(I_{max}(t)-I_{min}(t))}{(I_{max}(t)+I_{min}(t))} \tag{5}$$

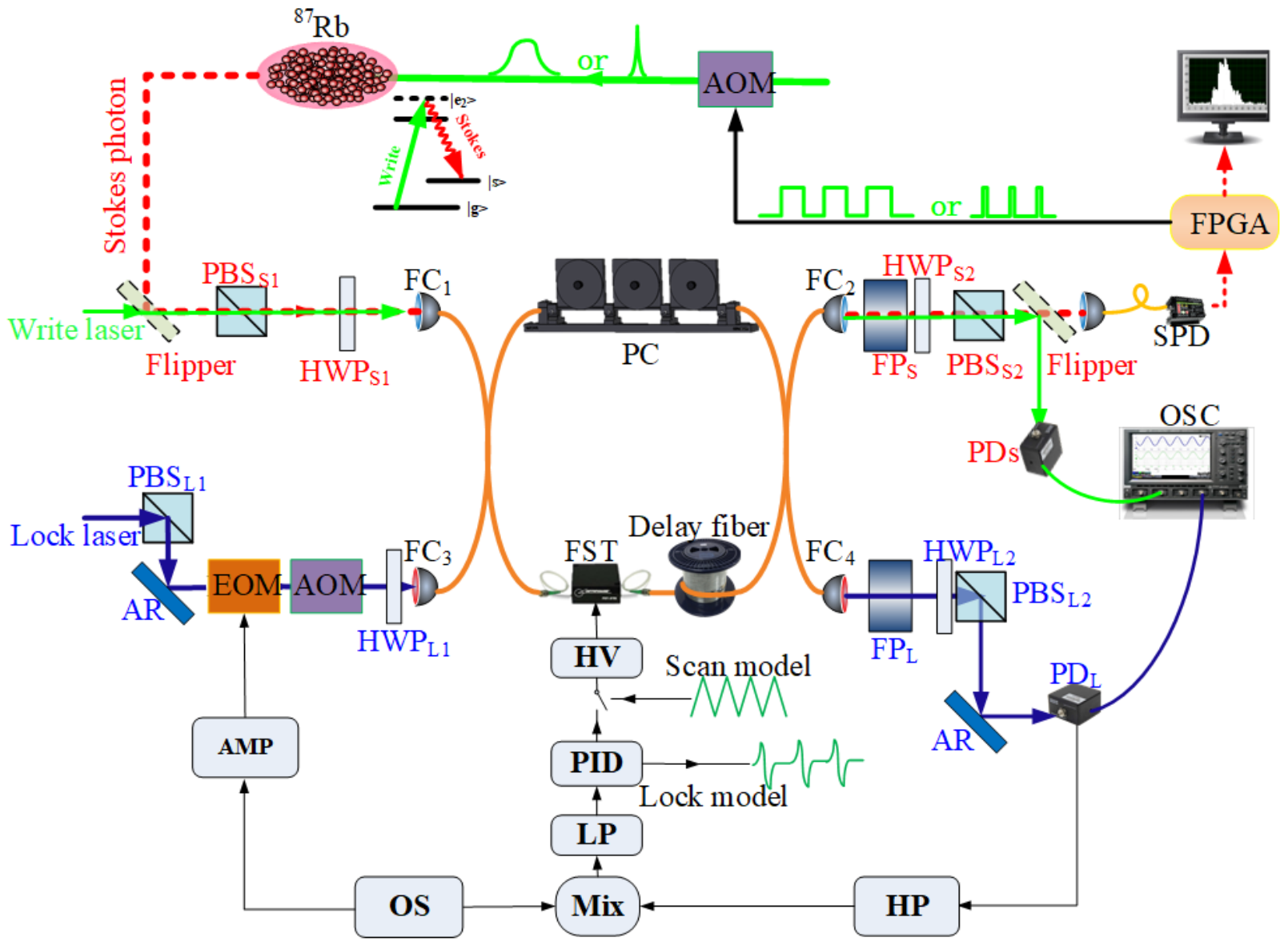

Fig. 1. Experimental Setup. FPs（FP$_L$）: Fabry-Perot filters, PBS: polarizing beam splitter; FPGA: programmable gate array; PBS: polarization-beam splitter; FC($_{1,2,3,4}$): fiber collimator; AOM: acousto-optic modulator; SMF: single mode optical fiber; HWP: half-wave plate; FST: Fiber Stretcher; PC: Polarization Controller; AMP: power amplifier; HV: High voltage amplifier; OS: oscillator; LP: lowpass; HP: high pass; PID: proportional-integral-derivative; OSC: oscilloscope; Flipper: Flip optical element adjustment frame; PDS(PDL): Write laser (lock laser) detector; SPD: single photon detector.

Flipper can switch the write laser to inject Stokes photons into the interferometer from $FC_1$ and control the interferometer length using the lock laser injected from $FC_3$. As shown in Fig.4, scanning the frequency of lock laser through AOM, the phase difference between Stokes photons and lock laser will change periodically, thus the counts of the photons in the interference region periodic fluctuate periodically.

The frequency difference between Stokes photon (write laser) and lock laser is ~6.8 GHz, in order to separate the Stokes photon (write laser) and the locked light at the position of $FC_1$ and $FC_2$, six Fabry-Perot filters (FP) are arranged in the detection path and the locking path respectively, $FP_L$ used for filtering write light or Stokes photon to extract locked light, $FP_S$ used for filtering locked light to extract write light or Stokes photon. Here the extinction ratio of the FP is the order of $10^{-16}$, and the transmittance of FP is 53%. In addition, polarization filtering are used for further separate Stokes photons and lock light, $PBS_{S1}$（$PBS_{S2}$） and $HWP_{S1}$（$HWP_{S2}$）used to ensure that horizontally（H） polarized Stokes photons are transmitted to the single photon detector in the H direction. $PBS_{L1}$（$PBS_{L2}$） and $HWP_{L1}$ （$HWP_{L2}$）used to ensure that vertical （V） polarized lock light are transmitted to the oscilloscope in the V direction.

## 4. Experimental Results

In the scan mode, Fig. 2 shows the interferograms of the delay fiber are 12 m and 212 m, blue line above (green line below) denotes the lock laser (write laser) signal. As the length difference $\Delta L(t)$ increases, the phase noise in $\varphi_L(t)(\varphi_S(t))$ is amplified based on equation (2), the signals in Fig.2(b) become fuzzy. Based on equation (3), the phase difference $\Delta\varphi(t)$ can change by controlling the frequency of AOM. Here, we list the in-phase case in Fig.2 with $\Delta\varphi(t)=2n\pi$ .

(a)

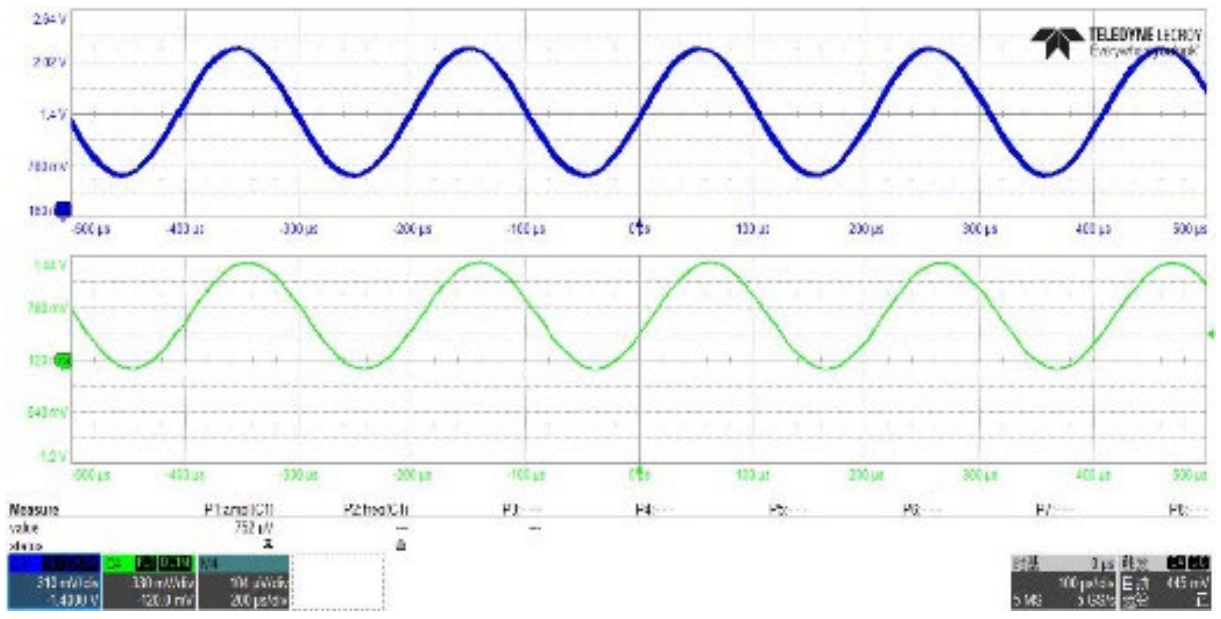

(b)

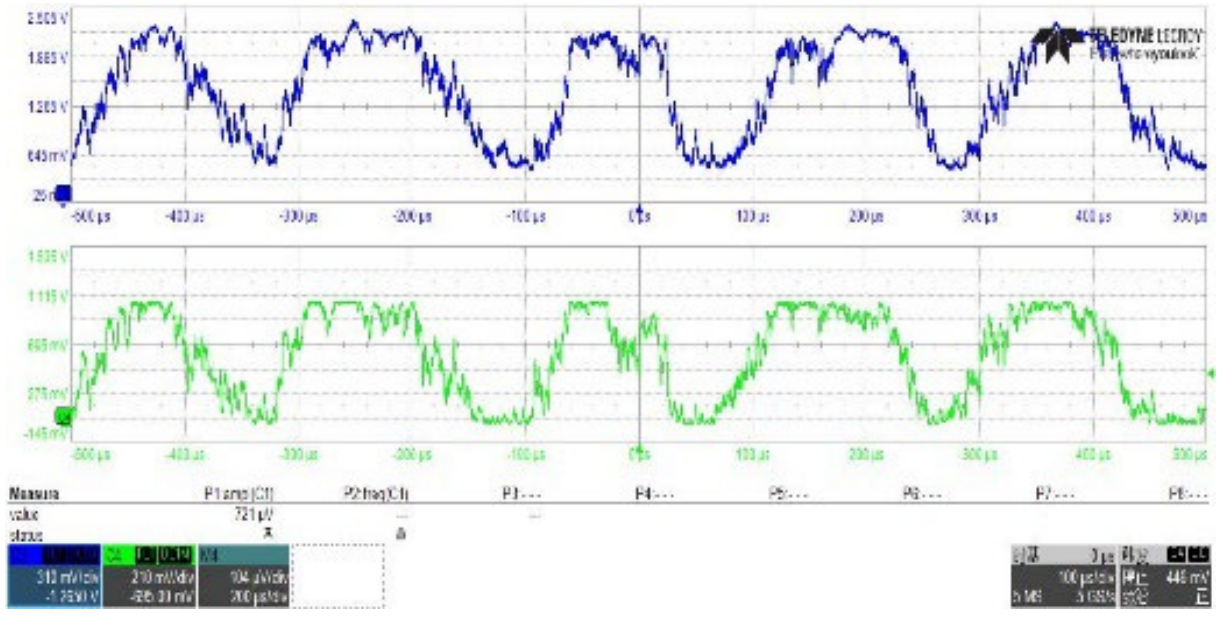

Fig.2. Interferograms collected by the oscilloscope of write laser and lock laser in the in-phase case with the delay fiber are (a)12 m and (b)212 m

When the lock laser (blue line above) is locked at the highest position, the write laser (green line below) is locked at the highest position (Fig. 3(a), (c)) in the in-phase case and the lowest position (Fig. 3(b), (d)) in the antiphase case. As the arm length difference of the interferometer increases, the phase noise is amplified, resulting in a coarse pattern curve. The intensity of the corresponding phase needs to be averaged over the observation time window, resulting in a significant reduction in the interference visibility. Based on equation (5), the corresponding interference visibility of different delay optical fibers are obtained, as shown in Fig.5 (a).

(a)

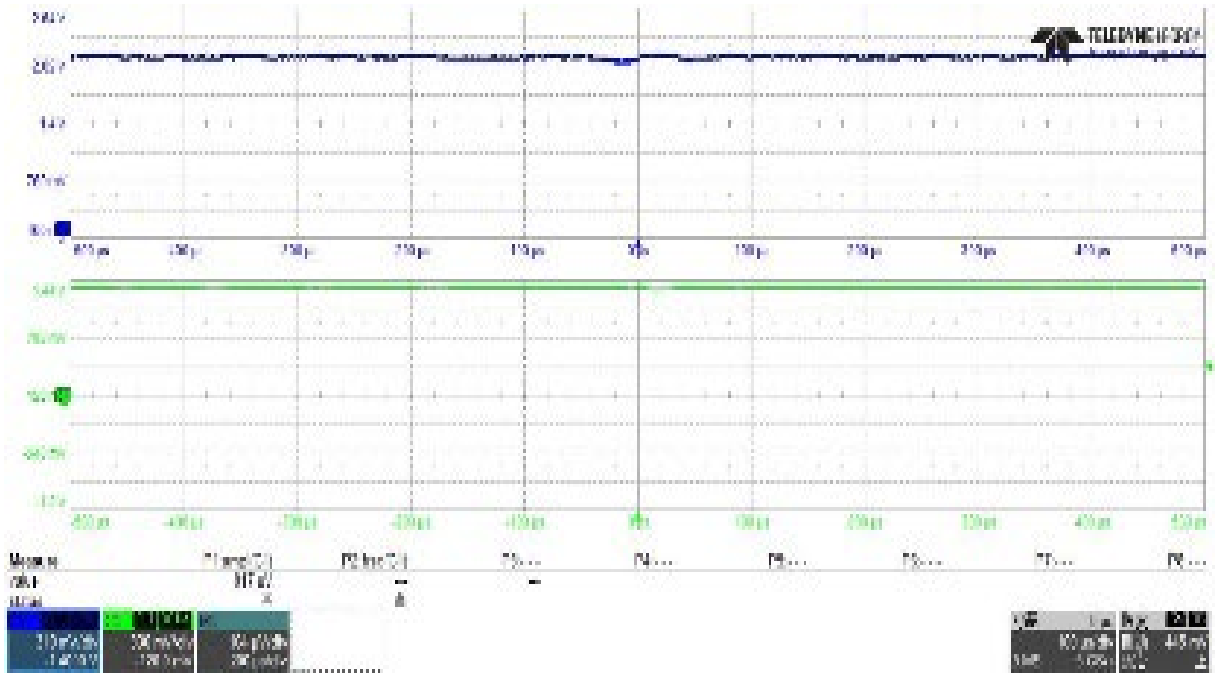

(b)

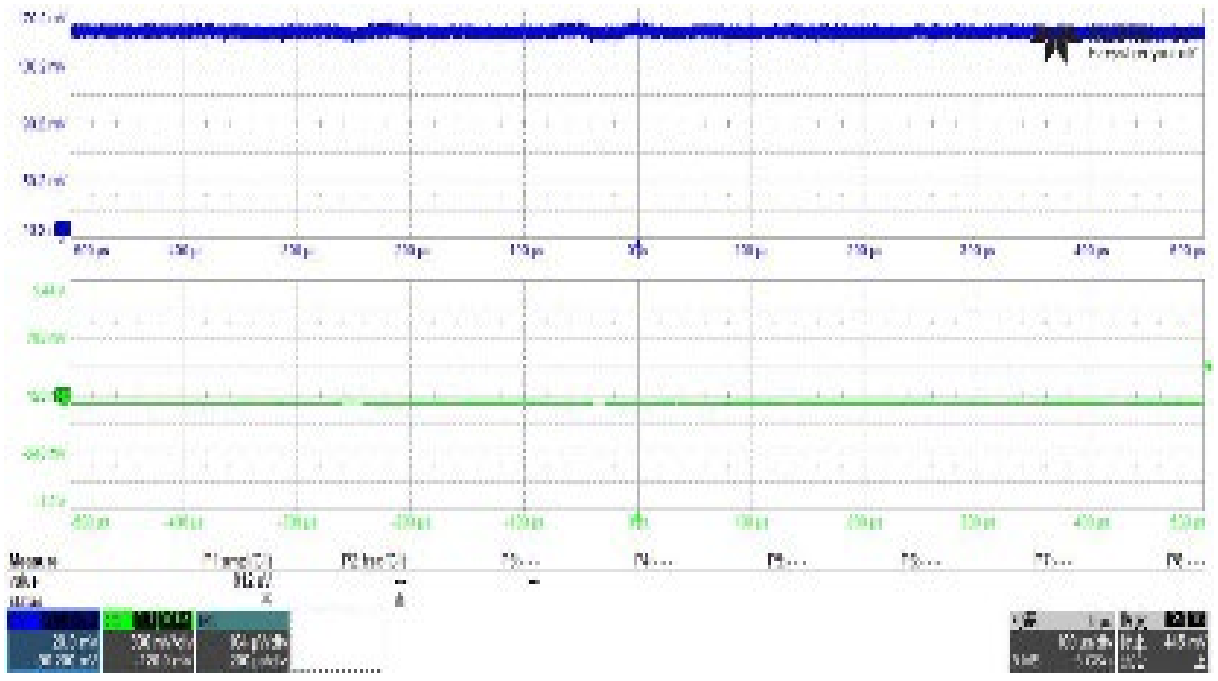

(c)

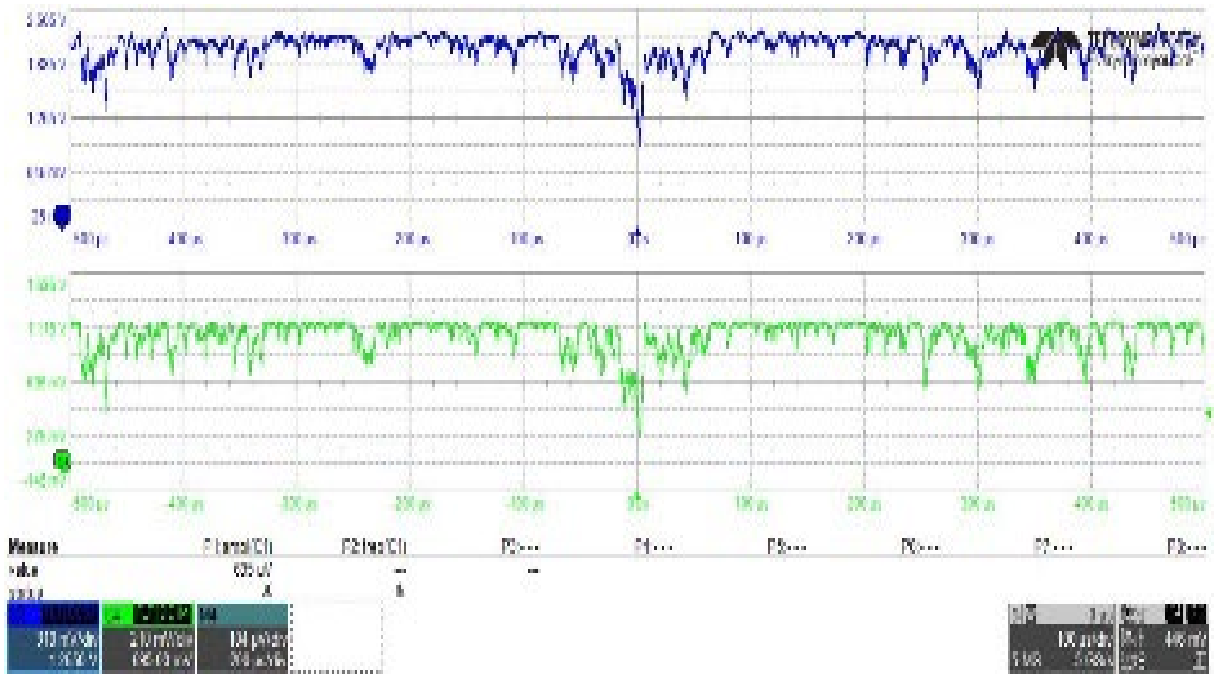

(d)

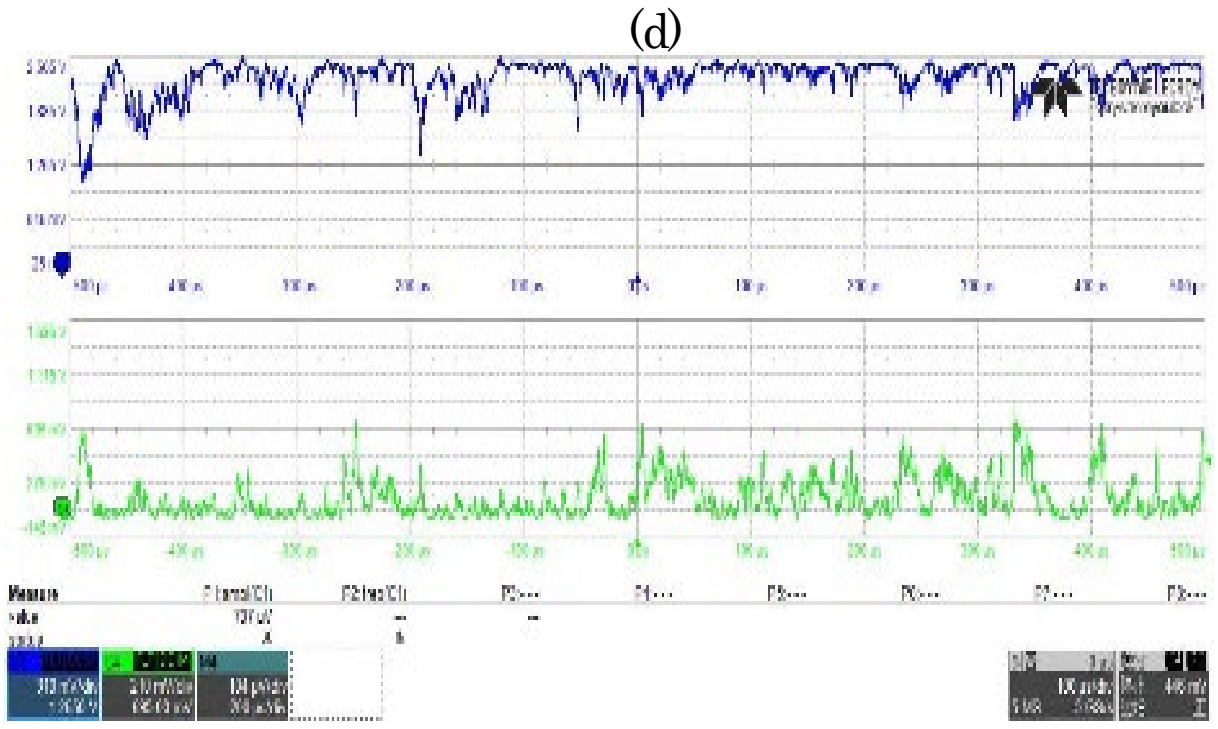

Fig.3. Signals collected by the oscilloscope of write laser and lock laser (a)(c) in the in-phase locked case and (b)(d) in the antiphase locked case with the delay fiber are (a)(b) 12m and (c)(d) 212m

Here, we list the photons counts in 90ns and 60ns interference region when 150ns and 1.06μs wave packet Stokes photons instead of the write laser injected into the interferometer as a function of frequency of AOM when the delay fiber is 12m and 212m as shown in Fig.4(a1) and Fig.4(b1). The interferograms have periods of ~18MHz and ~1MHz, respectively. Fig.4 (a2) and (a3) shows the highest and lowest counts distribution of photons at 197 MHz and 188 MHz of AOM when the delay fiber is 12m. Fig. 4 (b2) and (b3) shows the highest and lowest counts distribution of photons at 191.5MHz and 192 MHz of AOM when the delay fiber is 212m. The red circles are the measurements and the black squares are the denoising results. According to the black squares based on equation (5), the corresponding interference visibility is 93.5% and 66.7%, respectively. Due to the limitation of signal-noise ratio and laser frequency drift, we only measured the interference visibility of 150ns-1.06μs wave-packet-tunable photons.

(a1)

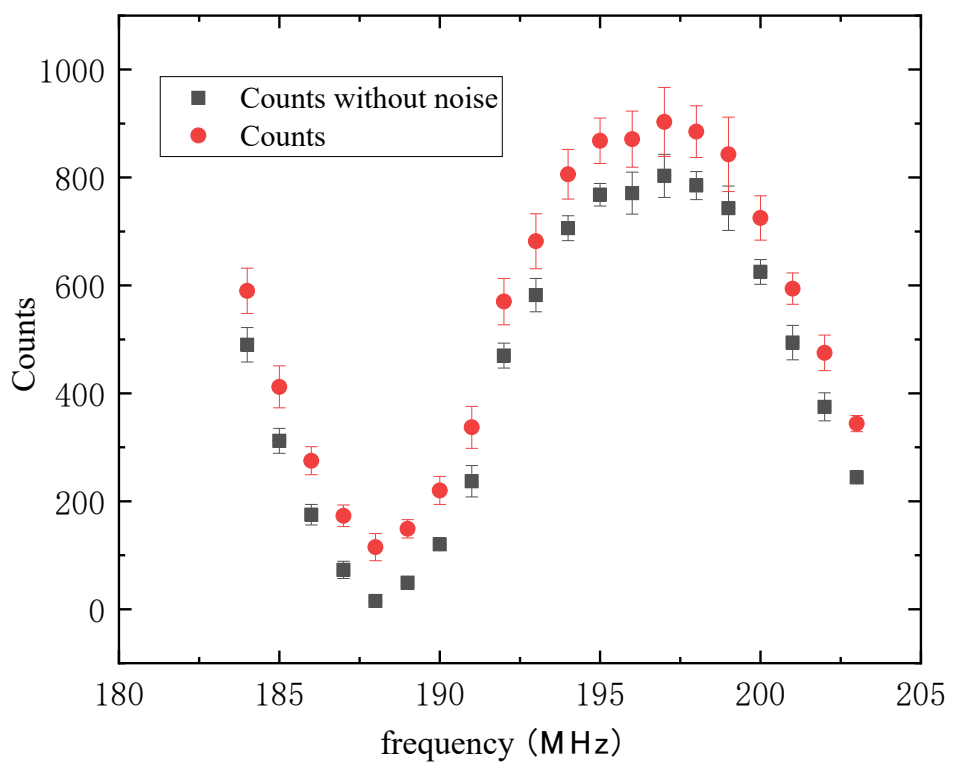

(b1)

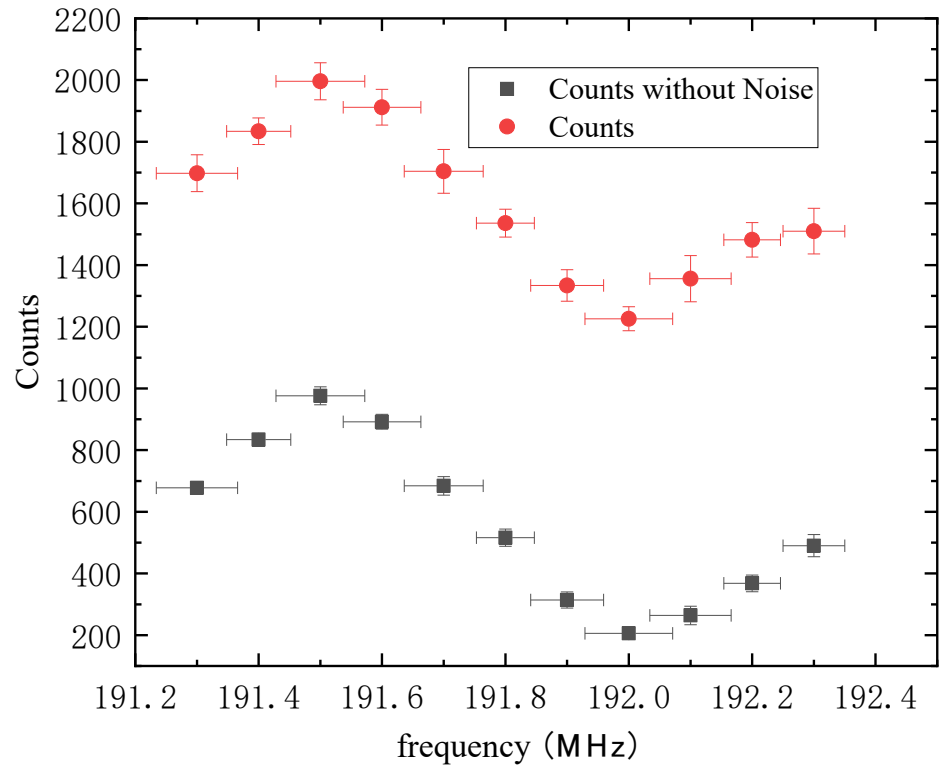

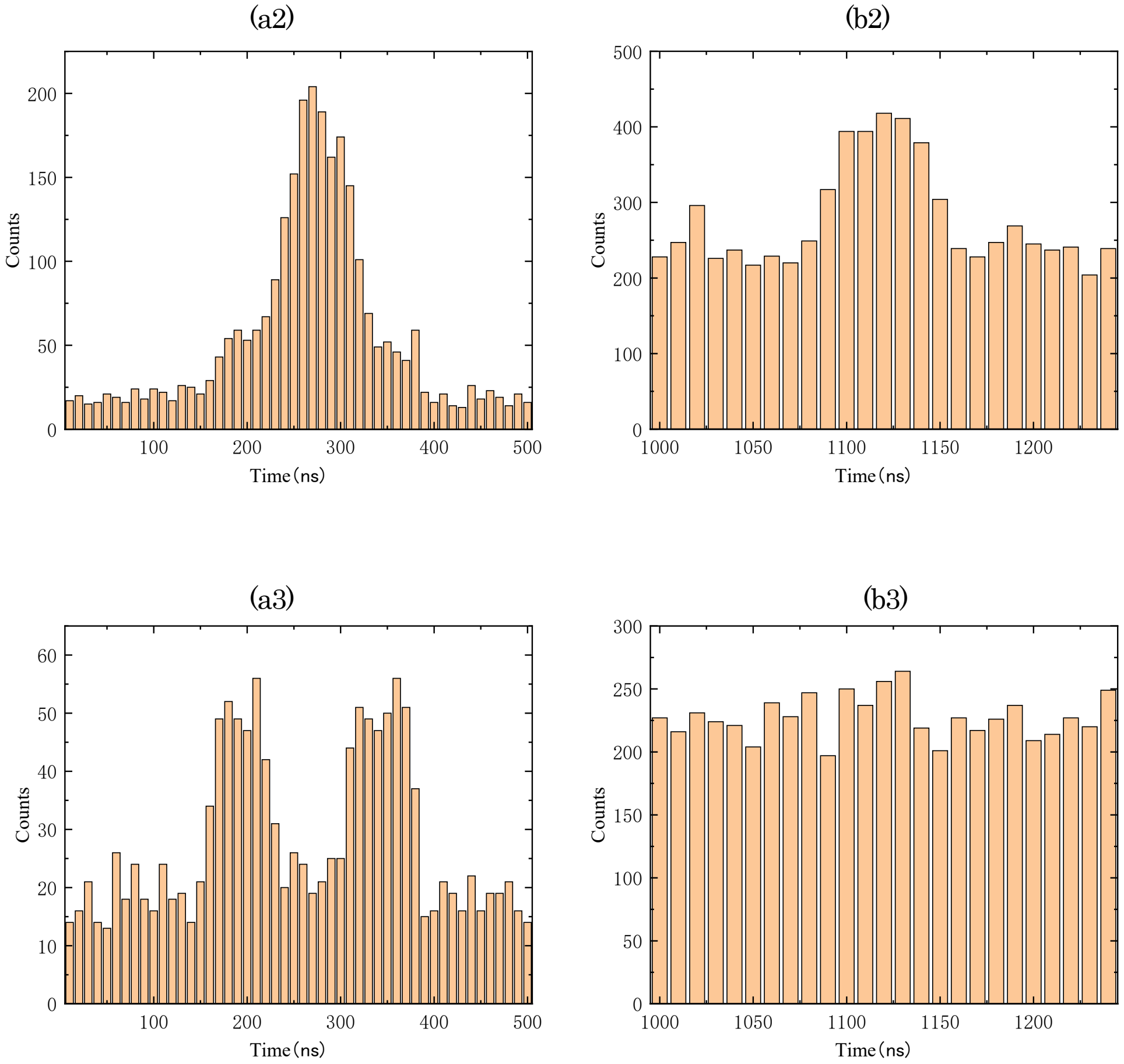

Fig. 4. The interferometry measurements of 150ns and 1.06us wave-packet photons. (a1) and (b1) shows the photons counts in 90ns and 60ns interference region, the red circles are the measured results, and the black squares are the denoising results. ;(a2) and (a3) shows the highest and lowest counts distribution of photons at 197MHz and 188 MHz of AOM when the delay fiber is 12m; (b2) and (b3) shows the highest and lowest counts distribution of photons at 191.5MHz and 192 MHz of AOM when the delay fiber is 212m. All detection events were binned into 10 ns.

We measured interference visibility V with eight different lengths 12m, 62m, 112m, 212m, 312m, 412m, 612m and 1012m of delay fibers. Based on equation (4), the red line is the fitting to the experimental data V of write laser in Fig. 5(a) with $\delta^2$=0.24, i.e., the bandwidth of the write laser is about 78KHz, and the coherence time is about 2.36μs when V=0.5. The blue line in Fig. 5(b) is the fitting to the experimental data V of wave-packet-tunable Stokes photons with $\delta^2$=0.96, i.e., the bandwidth of Stokes photons is 156KHz, and the coherence time is about 1.14μs when V=0.5.

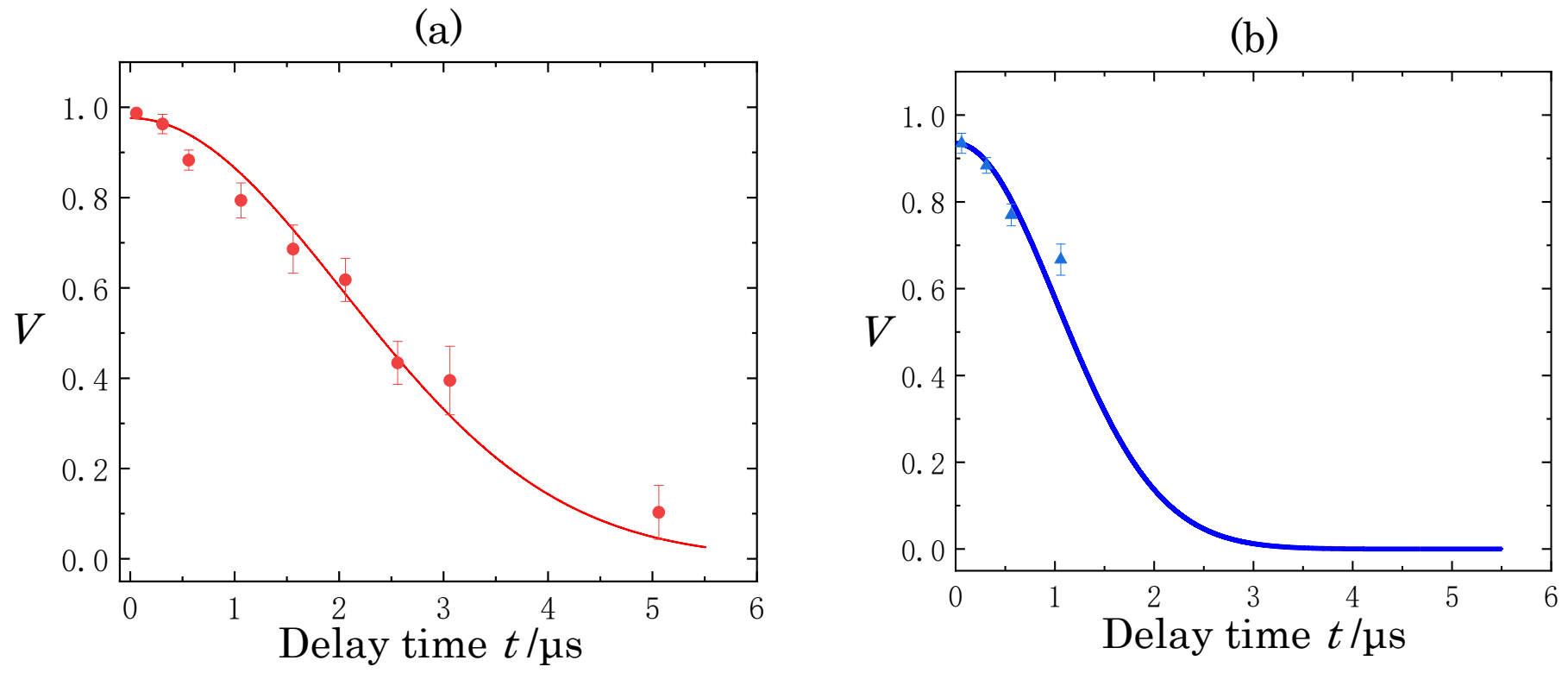

Fig.5. The measured interference visibility $V$ as a function of the delay time t. (a) The red solid curve is the fitting to the experimental data $V$ of write laser with $\delta^2$=0.24. (b) The blue solid curve is the fitting to the experimental data $V$ of wave-packet-tunable Stokes photons with $\delta^2$=0.96.

In order to measure the interference visibility for longer wave-packet photons, we will next use an ultra-stable cavity-locked laser to reduce the frequency drift of the write laser and then interact with the atoms to generate wave-packet Stokes photons with longer coherence times. We will also use a narrow bandwidth cavity to further suppress the noise and improve the signal-to-noise ratio, leading to higher measurement accuracy.

## 5. Conclusion

In this paper, the write laser and wave-packet-tunable Stokes photons are injected into the unequal arm interferometer. The coherence can be directly deduced from the decay of the visibility in the interferogram. Measuring the interference visibility of write laser and wave-packet-tunable Stokes photons with the corresponding length of the delay fiber, the results shows that the coherence time of the write laser is 2.36μs (bandwidth is 78KHz) can generate the Stokes photons with the coherence time is 1.14μs (bandwidth is 156KHz). The coherence measurement of wave-packet-tunable photons lays a foundation for the establishment of entanglement distribution between spatially separated memories in hybrid quantum network and the establishment of arbitrary length baseline telescope through wave-packet-tunable photon interference.

**Funding.** National Natural Science Foundation of China (62204232); the Fund for Shanxi Key Subjects Construction (1331); the Fundamental Research Program of Shanxi Province (202203021221118，2023030212218861); Research Project Supported by Shanxi Scholarship Council of China (20210038)

**Disclosures.** The authors declare no conflicts of interest.

**Data availability.** Data underlying the results presented in this paper are available from the corresponding authors upon reasonable request.

## References

1. H. J. Kimble, "The quantum internet," Nature **453**, 1023-1030 (2008).
2. S. Wehner, D. Elkouss, and R. Hanson, "Quantum internet: A vision for the road ahead," Science **362**, 1-9 (2018).
3. C. Simon, "Towards a global quantum network," Nat Photonics **11**, 678-680 (2017).
4. N. Sangouard, C. Simon, H. de Riedmatten, and N. Gisin, "Quantum repeaters based on atomic ensembles and linear optics," Reviews of Modern Physics **83**, 33-80 (2011).

5. L.-M. Duan, M. D. Lukin, J. I. Cirac, and P. Zoller, "Long-distance quantum communication with atomic ensembles and linear optics," Nature **414**, 413-418 (2001).
6. H.-J. Briegel, W. Dür, J. I. Cirac, and a. P. Zoller, "Quantum Repeaters: The Role of Imperfect Local Operations in Quantum Communication," Phys Rev Lett **81**, 5932-5935 (1998).
7. A. Kuzmich, W. P. Bowen, A. D. Boozer, A. Boca, C. W. Chou, L.-M. Duan, and H. J. Kimble, "Generation of Nonclassical Photon Pairs for Scalable Quantum Communication with Atomic Ensembles," Nature **425**, 731-734 (2003).
8. J. Laurat, K. S. Choi, H. Deng, C. W. Chou, and H. J. Kimble, "Heralded entanglement between atomic ensembles: preparation, decoherence, and scaling," Physical Review Letters **99**, 1-4 (2007).
9. B. Zhao, Y.-A. Chen, X.-H. Bao, T. Strassel, C.-S. Chuu, X.-M. Jin, J. Schmiedmayer, Z.-S. Yuan, S. Chen, and J.-W. Pan, "A millisecond quantum memory for scalable quantum networks," Nature Physics **5**, 95-99 (2008).
10. X.-H. Bao, A. Reingruber, P. Dietrich, J. Rui, A. Dück, T. Strassel, L. Li, N.-L. Liu, B. Zhao, and J.-W. Pan, "Efficient and long-lived quantum memory with cold atoms inside a ring cavity," Nature Physics **8**, 517-521 (2012).
11. S.-J. Yang, X.-J. Wang, X.-H. Bao, and J.-W. Pan, "An efficient quantum light–matter interface with sub-second lifetime," Nat Photonics **10**, 381-384 (2016).
12. X.-L. Pang, A.-L. Yang, J.-P. Dou, H. Li, C.-N. Zhang, E. Poem, D. J. Saunders, H. Tang, J. Nunn, I. A. Walmsley, and X.-M. Jin, "A hybrid quantum memory–enabled network at room temperature," Science Advances **6**, 1-8 (2020).
13. H. de Riedmatten, J. Laurat, C. W. Chou, E. W. Schomburg, D. Felinto, and H. J. Kimble, "Direct measurement of decoherence for entanglement between a photon and stored atomic excitation," Phys Rev Lett **97**, 1-4 (2006).
14. S. Chen, Y. A. Chen, B. Zhao, Z. S. Yuan, J. Schmiedmayer, and J. W. Pan, "Demonstration of a stable atom-photon entanglement source for quantum repeaters," Phys Rev Lett **99**, 1-4 (2007).
15. S. J. Yang, X. J. Wang, J. R. J. Li, X. H. Bao, and J. W. Pan, "Highly Retrievable Spin-Wave-Photon Entanglement Source," Physical Review Letters **114**, 1-5 (2015).
16. D.-S. Ding, W. Zhang, Z.-Y. Zhou, S. Shi, B.-S. Shi, and G.-C. Guo, "Raman quantum memory of photonic polarized entanglement," Nat Photonics **9**, 332-338 (2015).
17. Y. Wu, L. Tian, Z. Xu, W. Ge, L. Chen, S. Li, H. Yuan, Y. Wen, H. Wang, C. Xie, and K. Peng, "Simultaneous generation of two spin-wave–photon entangled states in an atomic ensemble," Physical Review A **93**, 1-8 (2016).
18. Y. Yu, F. Ma, X. Y. Luo, B. Jing, P. F. Sun, R. Z. Fang, C. W. Yang, H. Liu, M. Y. Zheng, X. P. Xie, W. J. Zhang, L. X. You, Z. Wang, T. Y. Chen, Q. Zhang, X. H. Bao, and J. W. Pan, "Entanglement of two quantum memories via fibres over dozens of kilometres," Nature **578**, 240-245 (2020).
19. Y. Wen, P. Zhou, Z. Xu, L. Yuan, H. Zhang, S. Wang, L. Tian, S. Li, and H. Wang, "Multiplexed spin-wave–photon entanglement source using temporal multimode memories and feedforward-controlled readout," Physical Review A **100**, 1-8 (2019).
20. S.-Z. Wang, M.-J. Wang, Y.-F. Wen, Z.-X. Xu, T.-F. Ma, S.-J. Li, and H. Wang, "Long-lived and multiplexed atom-photon entanglement interface with feed-forwardcontrolled readouts," Communications Physics **4**, 1-9 (2021).
21. C. Clausen, I. Usmani, F. Bussieres, N. Sangouard, M. Afzelius, H. de Riedmatten, and N. Gisin, "Quantum storage of photonic entanglement in a crystal," Nature **469**, 508-511 (2011).
22. P. Farrera, G. Heinze, B. Albrecht, M. Ho, M. Chavez, C. Teo, N. Sangouard, and H. de Riedmatten, "Generation of single photons with highly tunable wave shape from a cold atomic ensemble," Nature Communications **7**, 1-6 (2016).
23. M. Almendros, J. Huwer, N. Piro, F. Rohde, C. Schuck, M. Hennrich, F. Dubin, and J. Eschner, "Bandwidth-tunable single-photon source in an ion-trap quantum network," Phys Rev Lett **103**, 1-4 (2009).
24. G. B. Xavier and J. P. v. d. Weid, "Stable single-photon interference in a 1 km fiber-optic Mach–Zehnder interferometer with continuous phase adjustment," Optics Letters **36**(2011).
25. S.-B. Cho and T.-G. Noh, "Stabilization of a long-armed fiber-optic singlephoton interferometer," OPTICS EXPRESS **17**, 19027-19032 (2009).
26. F. Jelezko, A. Volkmer, I. Popa, K. K. Rebane, and J. Wrachtrup, "Coherence length of photons from a single quantum system," Physical Review A **67**, 1-4 (2003).
27. D. Gottesman, T. Jennewein, and S. Croke, "Longer-baseline telescopes using quantum repeaters," Phys Rev Lett **109**, 070503 (2012).

28. Y. Li, Y. Wen, S. Wang, C. Liu, H. Liu, M. Wang, C. Sun, Y. Gao, S. Li, and H. Wang, "Generation of entanglement between a highly wave-packet-tunable photon and a spin-wave memory in cold atoms," Opt Express **30**, 2792-2802 (2022).
29. N. Maring, P. Farrera, K. Kutluer, M. Mazzera, G. Heinze, and H. de Riedmatten, "Photonic quantum state transfer between a cold atomic gas and a crystal," Nature **551**, 485-488 (2017).
30. J. Minář, H. de Riedmatten, C. Simon, H. Zbinden, and N. Gisin, "Phase-noise measurements in long-fiber interferometers for quantum-repeater applications," Physical Review A **77**, 1-8 (2008).